# Absence of near-ambient superconductivity in LuH$_{2\pm x}$N$_y$

Xue Ming[†], Ying-Jie Zhang[†], Xiyu Zhu[*], Qing Li[*], Chengping He, Yuecong Liu, Bo Zheng, Huan Yang, and Hai-Hu Wen[*]

National Laboratory of Solid State Microstructures and Department of Physics, Collaborative Innovation Center of Advanced Microstructures, Nanjing University, Nanjing 210093, China

[†]These authors contributed equally to this work. [*]e-mail: hhwen@nju.edu.cn; liqing1118@nju.edu.cn; zhuxiyu@nju.edu.cn

**Recently near-ambient superconductivity was claimed in N-doped lutetium hydride (ref. [1]). This induces a worldwide fanaticism about the dream of room temperature superconductivity under low pressures. By using a high pressure and high temperature synthesis technique, we have successfully obtained the nitrogen doped lutetium hydride (LuH$_{2\pm x}$N$_y$) with a dark-bluish color and a structure with the space group of $Fm\overline{3}m$ evidenced by x-ray diffraction. This structure is the same as that reported in ref. [1]. The energy dispersive X-ray spectroscopy (EDS) confirmed the existence of nitrogen in some areas of the samples. At ambient pressure, we witness a kink of resistivity and magnetization at about 300 K, which may correspond to a rearrangement of hydrogen/nitrogen atoms, namely a structural transition. However, by applying a pressure from 1 GPa to 6 GPa, we have seen a progressively optimized metallic behavior without showing superconductivity down to 10 K. Temperature dependence of magnetization shows**

**a roughly flat feature between 100 and 320 K, and the magnetization increases with magnetic field at 100 K, all these are not expected for superconductivity at 100 K. Thus, we conclude the absence of near-ambient superconductivity in this nitrogen-doped lutetium hydride under pressures below 6 GPa.**

Metallic hydrogen and hydrogen-rich materials provide interesting platforms for searching room temperature superconductivity since it was proposed theoretically by Ashcroft *et al.*[2]. However, due to the extreme high pressure required, experimentally it is difficult to achieve high temperature superconductivity[3,4]. Then theorists proposed that the polyhydrides have the potential to realize high temperature superconductivity due to the effect of internal chemical pressure[5]. Later, it was indeed verified, as the discovery of high temperature superconductivity in $H_3S$ with a superconducting transition temperature ($T_c$) above 200 K under a high pressure (~200 GPa) both in theory and experiment[6-8]. After that, more and more hydrogen-rich high temperature superconductors have been discovered, such as $LaH_{10}$[9,10], $CaH_6$, *etc.*[11-15]. However, according to the basic understanding of Bardeen-Cooper-Schrieffer (BCS) theory, superconductivity would rely on strong electron-phonon coupling with high Debye temperature. According to the McMillan formula, if we assume a Debye temperature of 500 K, the Coulomb screening constant $\mu^* = 0.13$, we find that the electron-phonon coupling constant λ would be as large as 12.2 when the material has a $T_c$ of 100 K. This huge λ cannot allow a stable lattice structure, thus this high temperature superconductivity can only be achieved in such systems when they are protected by extremely high pressures.

Recently, superconductivity at about 294 K in N-doped Lu hydride under only 1 GPa was reported [1], which is extremely interesting and important if the observation of superconductivity could be repeated. As reported by Dias and his coworkers, the dark-bluish ternary compound (as the formula $LuH_{3-\delta}N_{\varepsilon}$ used by them) can be tuned to a near-ambient superconductor by a relatively low pressure (1-2 GPa), accompanied with a color change from blue to pink and red. And in the main-text of their article, superconductivity was claimed by the measurements of several quantities, including resistivity, magnetization and specific heat. While, it seems that this conclusion was drawn mainly based on the treatment of subtracting backgrounds from the raw signals, otherwise it would give no trace of the zero resistance state and the perfect diamagnetism expected for a superconductor. Actually, in previous experiments, superconductivity with much lower transition temperatures were reported by other groups under high pressures in Lu hydride[16, 17], which is different from that presented in ref. [1]. For example, $LuH_3$ ($Fm\bar{3}m$) shows superconductivity at about 12.4 K under the pressure of about 122 GPa[16], while superconductivity in $Lu_4H_{23}$ ($Pm\bar{3}n$) can exist at temperatures up to about 71 K at 218 GPa[17]. Thus, comparing with above results of Lu hydrides, the discovery of near-ambient superconductivity in N-doped lutetium hydride is really striking. It generates great curiosity whether room temperature superconductivity really exists in this N-doped lutetium hydride under relatively low pressures.

In this manuscript, we report the synthesis of the similar N-doped Lu hydrides, and also the experimental results on these samples. By using a high pressure and high temperature synthesis technique, we successfully obtained metallic samples with dark-blue color. The structure and composition analysis confirmed the phase of $LuH_{2\pm x}N_y$ (with a space group: $Fm\overline{3}m$), which is consistent with the result reported by Dias and coworkers[1]. Unfortunately, our results show the absence of superconductivity down to 10 K in the N-doped Lu hydride prepared by our group.

**Sample characterization and physical properties at ambient pressure.**

The inset of Fig. 1a presents the picture of our samples with dark-bluish color. And Fig. 1a shows the X-ray diffraction (XRD) pattern and Rietveld refinement for the sample $LuH_{2\pm x}N_y$. As we can see, the experimental data can be well fitted by the structure of $LuH_2$ with the space group of $Fm\overline{3}m$ and lattice parameter $a = 5.032$ Å. In order to compare our sample with the one reported in ref. [1], we downloaded the raw XRD data from ref. [1] and draw it together with our data after normalization. As shown in Fig. 1b, the two sets of XRD data with peaks almost coincide, indicating the same structure. We notice that this structure may be indexed to the similar one of $LuH_2$. The inset of Fig. 1b shows an enlarged view of the X-ray diffraction pattern with $2\theta$ from 29° to 37°. As we can see, the peaks of the two samples coincide very well with each other. Concerning impurities, comparing with that in the ref. [1], actually our samples have much less amount of impurities, nor the phase of $Lu_2O_3$. Therefore, according to our XRD data, we obtained the compound almost the same as that reported in ref. [1]. Figure 1c displays the SEM images of our sample $LuH_{2\pm x}N_y$. Energy dispersive X-ray spectroscopy (EDS) is used to analyze the element composition. We measured 10 spots randomly marked by the black crosses as shown in the inset of Fig. 1c. The typical EDS of spot 1 is shown in Fig. 1c. From the spectrum, we can see a weak peak from nitrogen. And as shown in Fig. 1d, nitrogen can be detected at 4 of the 10 spots. Among them, spot 1 has the highest nitrogen content, which can reach 1.38% by weight (including lutetium and oxygen). The oxygen here should not be an intrinsic composition of the sample since it can be easily absorbed when the sample is loaded in air before the SEM/EDS measurement. However, we must point out that, at 6 out of the 10 points, we have not detected nitrogen. Regarding the insensitivity of the EDS technique to nitrogen, this spatial fluctuation may be understandable, but at least it indicates that some amount of nitrogen exists in the samples. Since it is impossible to detect the hydrogen atoms by EDS, while XRD shows that the structure is quite consistent with $LuH_2$, we define the formula of our samples as $LuH_{2\pm x}N_y$.

Figure 2a displays the temperature dependence of resistivity for $LuH_{2\pm x}N_y$ under ambient pressure. Two samples were selected and measured to ensure the reliability of the results. As we can see, the resistivity of both samples shows a metallic behavior. The $\rho$-$T$ curves are roughly linear from 40 K to 300 K and exhibit a quadratic temperature dependence at lower temperatures, indicating a Fermi liquid behavior. In addition, we notice that a kink of resistivity exists at about 316 K for sample 1, which is consistent with the separation of ZFC and FC curves shown in Fig. 2b. We think that this may be related to some kind of transition, such as the temperature driven rearrangement of H/N atoms. Figure 2b presents the temperature dependence of magnetic moment in zero-field cooling (ZFC) and field-cooling (FC) modes at 10 Oe for the prepared $LuH_{2\pm x}N_y$ sample under ambient pressure. The temperature dependent magnetic moment shows a paramagnetic behavior. Interestingly, the data of ZFC and FC start to separate at a temperature of about 315 K, which may also indicate some unknown transition which needs to be further explored. We must emphasize that this signal is very weak, being in the scale of the systematic error of the instrument.

**Temperature-dependent electrical resistance under high pressures.**

To explore possible near-ambient superconductivity in $LuH_{2\pm x}N_y$ under high pressure, we measured $R$(T) curves of the dark-bluish $LuH_{2\pm x}N_y$ sample by using a DAC in an extended pressure range up to 6 GPa. Figure 3a shows the temperature dependence of electrical resistance from 10 K to 300 K under various pressures. The resistance at room temperature progressively decreases with the increase of pressure up to 6.3 GPa. Under low pressures, the $R$(T) curve shows a hump structure around 300 K, and such feature then becomes gradually weakened with the increase of pressure. To verify whether the decrease in resistance around room temperature is related to a possible superconducting transition, we have also measured $R$(T) curves of $LuH_{2\pm x}N_y$ under various magnetic fields under a pressure of 1.6 GPa. As we can see from Fig. 3b, the resistance under the magnetic field exhibits a similar temperature dependence which does not show any drifting to lower temperatures as expected for a superconductor when a magnetic field is applied. This general behavior is completely inconsistent with that of superconducting materials under magnetic fields. We also notice that the $R$(T) curves in Fig. 3a show a clear upturn in low temperature region, which may be related to strong scattering by the grain boundaries of the polycrystalline sample. Therefore, another experiment (run 2) was performed with the original sample but grounded into powder, and later these powder are pressed and filled in a DAC, the measured resistivity data are shown in Fig. 3c. We can see that the resistance upturn disappears but an even larger residual resistance appears at low temperatures, and the sample exhibits typical Fermi liquid behavior at low temperatures. However, we still do not observe a superconducting related transition or zero-resistance state of $LuH_{2\pm x}N_y$ in our electrical resistance measurement down to 10 K with pressures ranging from 1.3 to 6.8 GPa.

**Temperature- and Field-dependent magnetization under high pressures.**

To prove whether there is a diamagnetic signal due to the Meissner effect of the possible superconductivity in the as-grown samples, we measured the dc magnetization $M$(T) of $LuH_{2\pm x}N_y$ in the temperature region of 100 to 320 K with the pressure of 1 GPa and 2.1 GPa, respectively. The temperature dependence of magnetic moment curves at 60 Oe in the ZFC and FC mode with different pressures are shown in Fig. 4 a and b. Under these two pressures, the magnetization with a negative value decreases with decreasing temperature and does not show a sudden drop behavior in the magnetization curve. This negative value of magnetization is not an intrinsic feature of the sample, but due to the background signal of the DAC device (HMD cell, see **Methods**). With the same setup, we have detected a clear superconducting transition in Bi samples under pressures [18]. In order to get the magnetization signal purely from the sample, we also measured the background signal of the HMD cell at the same field (60 Oe) and the same pressures. The data of background signal are presented in Extended Fig.1. The net signal after removing the related background is shown in the insets of Fig. 4a and b. One can see that the net signal of magnetization is positive with a roughly flat feature in the temperature region from 100 K to 320 K. Figure 4 c displays the isothermal magnetization $M$(H) curves for $LuH_{2\pm x}N_y$ at 100 K under pressures of 1 GPa (open square) and 2.1 GPa (open circle). The $M$(H) curve shows a rough linear behavior with a negative slope from 0 Oe to 6000 Oe. This is not a feature of the Meissner effect, but is due to the background again. To prove that, we have measured one $M$(H) curve at 320 K under 2.1 GPa (up triangle), and one curve at 100 K for the empty HMD cell (solid square). In Fig. 4d, we show the net $M$(H) curves after subtracting related backgrounds. It is clear that all net $M$(H) curves exhibit a rough linear behavior with a positive correlation. This corresponds to a possible paramagnetic behavior. Our magnetization measurements with the data of either temperature dependence or the isothermal magnetization curves $M$(H) all show that there is no any trace of superconductivity at and above 100 K.

**Sample color under high pressures.**

One of the surprising phenomena in Dias and coworker's article is the color change of the samples with increasing pressure. Here we also show the colors of our nitrogen doped samples under different pressures in Fig. 5. Surprisingly, the dark-bluish color of the $LuH_{2\pm x}N_y$ sample was maintained up to 5.2 GPa, and the similar color is still maintained after the pressure is released. Considering the higher purity of the $LuH_{2\pm x}N_y$ phase in our samples (inferred from the XRD data), the pressure-induced color changes in ref. [1] seem unlikely to come from the main phase of $LuH_{2\pm x}N_y$. One possibility to interpret the discrepancy is that the concentrations of nitrogen are different in our samples and theirs, since the synthesis methods are quite different. We have also witnessed a color change from dark-bluish at ambient pressure to pink and red under pressures of 2.5 and 5.0 GPa in the sample $LuH_2$ (see Extended Fig.2). Thus the threshold of this color change may shift to higher pressures with nitrogen doping. It deserves to be further explored how the nitrogen doping influences the color change in this interesting system. Despite the unwitnessed color change induced by pressure in our nitrogen doped samples up to 5.2 GPa, the temperature dependence of resistivity in our samples and that in ref.[1] in the so-called superconducting sample all show a similar metallic behavior down to 10 K without showing superconductivity.

In summary, we have successfully synthesized the dark-blue N-doped lutetium hydrides $LuH_{2\pm x}N_y$. The XRD confirmed that our sample is exactly the same as the main phase reported in ref. [1]. Meanwhile, the existence of nitrogen in our samples has been confirmed by EDS analysis. However, the color change reported in ref. [1] under high pressure has not been observed. We argue that this color change may be induced more easily in pure $LuH_2$ and the threshold pressure will be enhanced by nitrogen doping concentration. Our experiments clearly reveal that no superconductivity exists in $LuH_{2\pm x}N_y$ from ambient pressure to 6.3 GPa with temperatures down to 10 K.

**Note:** During the preparation of this manuscript, we became aware of a report of the color change and absence of superconductivity in LuH$_2$ under pressures [19].

**Methods**

**Sample preparation.** We synthesized polycrystalline samples of LuH$_{2\pm x}$N$_y$ using piston-cylinder type high pressure apparatus (LP 1000-540/50, Max Voggenreiter). The NH$_4$Cl and excessive CaH$_2$ were used as the source of nitrogen and hydrogen, according to the chemical equation written as 2NH$_4$Cl + CaH$_2$→ CaCl$_2$ + 2NH$_3$ +H$_2$. The NH$_4$Cl (Alfa Aesar 99.99%) was mixed well with CaH$_2$ (Alfa Aesar 98%) in a molar ratio of 2:8 and pressed into a tablet. Then, the tablet and pellet made by Lu pieces with silver color were separated by a BN pellet and sealed into a gold capsule. Then the gold capsule was placed in a BN capsule and heated at 300°C for 10 hours under 2 GPa. Finally, we obtained the LuH$_{2\pm x}$N$_y$ with a dark blue color.

The X-ray diffraction (XRD) measurements were performed on a Bruker *D8* Advanced diffractometer with the Cu$K_{\alpha 1}$ radiation. The Rietveld refinements were done by using the software of *TOPAS4.2* [20]. The scanning electron microscope (SEM) photograph and the energy dispersive X-ray microanalysis spectrum were obtained by Phenom ProX (Phenom) at an accelerating voltage of 15 kV.

Temperature dependent resistivity measurements under ambient and high pressure were carried out with a physical property measurement system (PPMS-9T, Quantum Design). The high pressure was generated by a Diamond anvil cells (DAC) made of BeCu alloy with two opposing anvils. A four-probe van der Pauw method with platinum foil as electrodes was applied for resistance measurements. dc magnetization measurements were performed with a SQUID-VSM-7T (Quantum Design). The dc magnetization measurements at high pressures were accomplished by using the DAC (attachment to a PPMS) designed by the Honest Machinery Designer's office (HMD). The gasket is made by BeCu. The anvils with beveled culet size of 400 μm and 600 μm culets were used to generate high pressures. NaCl and Daphne 7373 were used as the pressure transmitting medium during the resistive and magnetic susceptibility measurements, respectively. The pressure was measured at room temperature using the ruby fluorescence method [21].

**Data availability**

All data needed to evaluate the conclusions in the paper are present in the paper. Additional data related to this paper may be requested from the authors.

**Acknowledgements**

This work was supported by the National Key R&D Program of China (No. 2022YFA1403201), National Natural Science Foundation of China (Nos. 12061131001,

**Author contributions**

The samples and SEM/EDX analysis were made by X.M., X.Y.Z., C.P.H., B.Z. and H.H.W. The XRD data were collected by Y.C.L. and X.M. The resistivity and magnetization were measured by Q.L. and Y.J.Z. The photos were taken by H.Y. All authors joined the analysis and agreed to publish the data. The manuscript was written by Q.L., X.Y.Z. and H.H.W. H.H.W. conceived and supervised the whole study.

**Competing interests**

The authors declare that they have no competing interests.

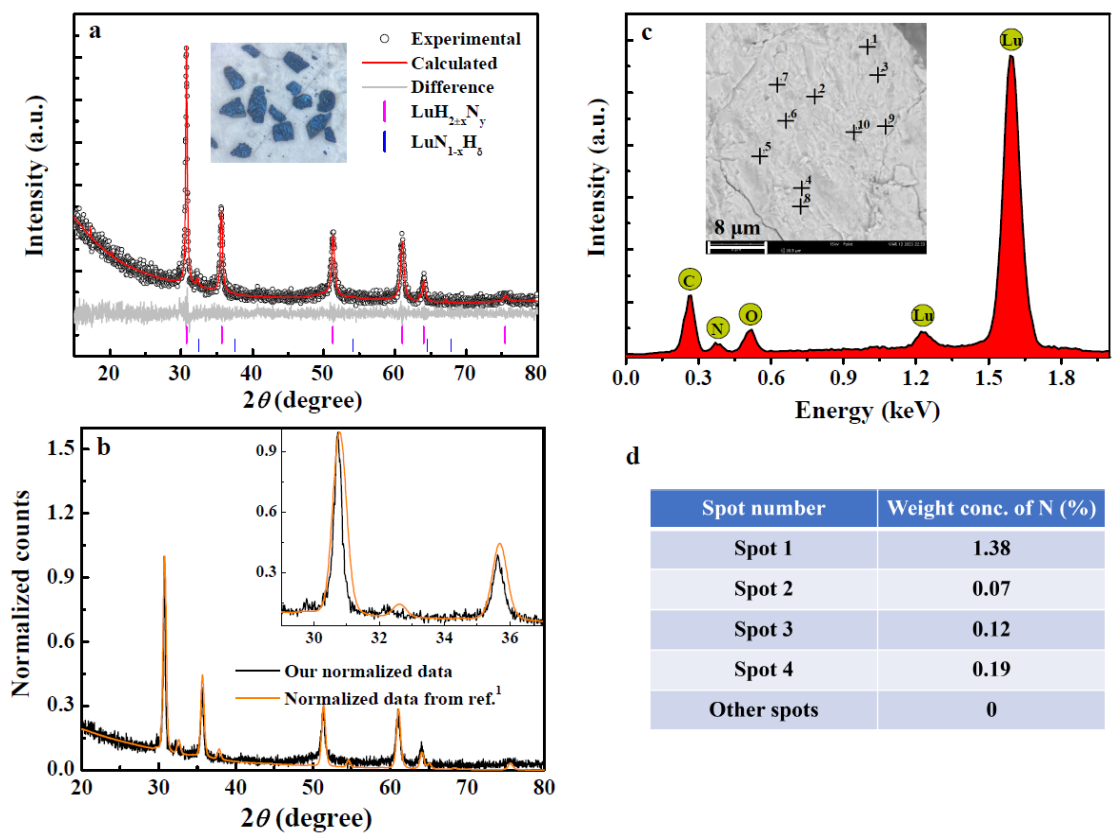

**Fig. 1 | XRD, Rietveld refinement, SEM images and EDS analyses for LuH$_{2\pm x}$N$_y$. a** Powder X-ray diffraction patterns of the LuH$_{2\pm x}$N$_y$ and Rietveld fitting curves (red lines) to the data. The inset shows the picture of LuH$_{2\pm x}$N$_y$ which exhibits a dark-bluish color. **b** The XRD comparison between ours and that downloaded from ref. [1]. **c** SEM images and typical energy dispersive spectroscopy (EDS) of spot 1. **d** A table for the weight concentration of N (%) by EDS analysis at 10 spots of the sample.

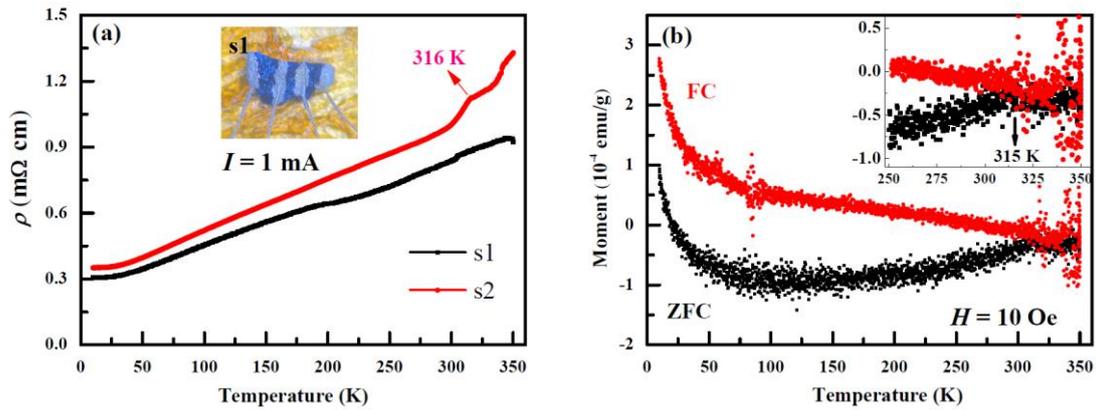

**Fig. 2 | Temperature-dependence of resistivity and magnetic moment for LuH$_{2\pm x}$N$_y$ under ambient pressure. a** Temperature dependence of resistivity for the LuH$_{2\pm x}$N$_y$ samples under ambient pressure. The inset shows the image of the measured sample (s1) with electrodes attached at ambient pressure. **b** Temperature dependence of magnetic moment for the prepared LuH$_{2\pm x}$N$_y$ samples in both ZFC and FC modes at 10 Oe under ambient pressure. The inset shows the data between 250 K and 350 K.

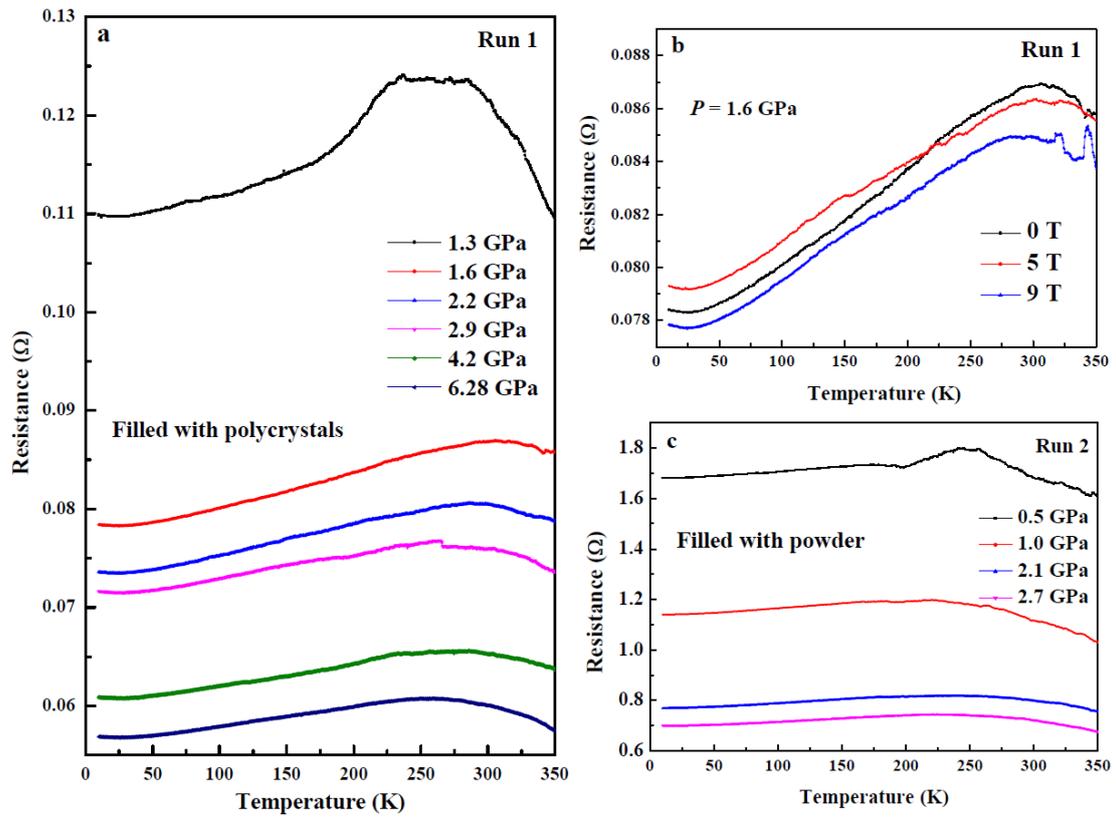

**Fig. 3 | Temperature-dependent electrical resistance for LuH$_{2\pm x}$N$_y$ at different pressures. a** Temperature dependence of the electrical resistance of LuH$_{2\pm x}$N$_y$ from 10 to 350 K with pressures up to 6.3 GPa (DAC filled with polycrystals). **b** Temperature dependence of the electrical resistance of LuH$_{2\pm x}$N$_y$ measured at different magnetic fields up to 9 T at 1.6 GPa (DAC filled with polycrystals). **c** Temperature dependence of the electrical resistance of LuH$_{2\pm x}$N$_y$ up to 2.7 GPa for another run with the DAC filled with powder of the sample.

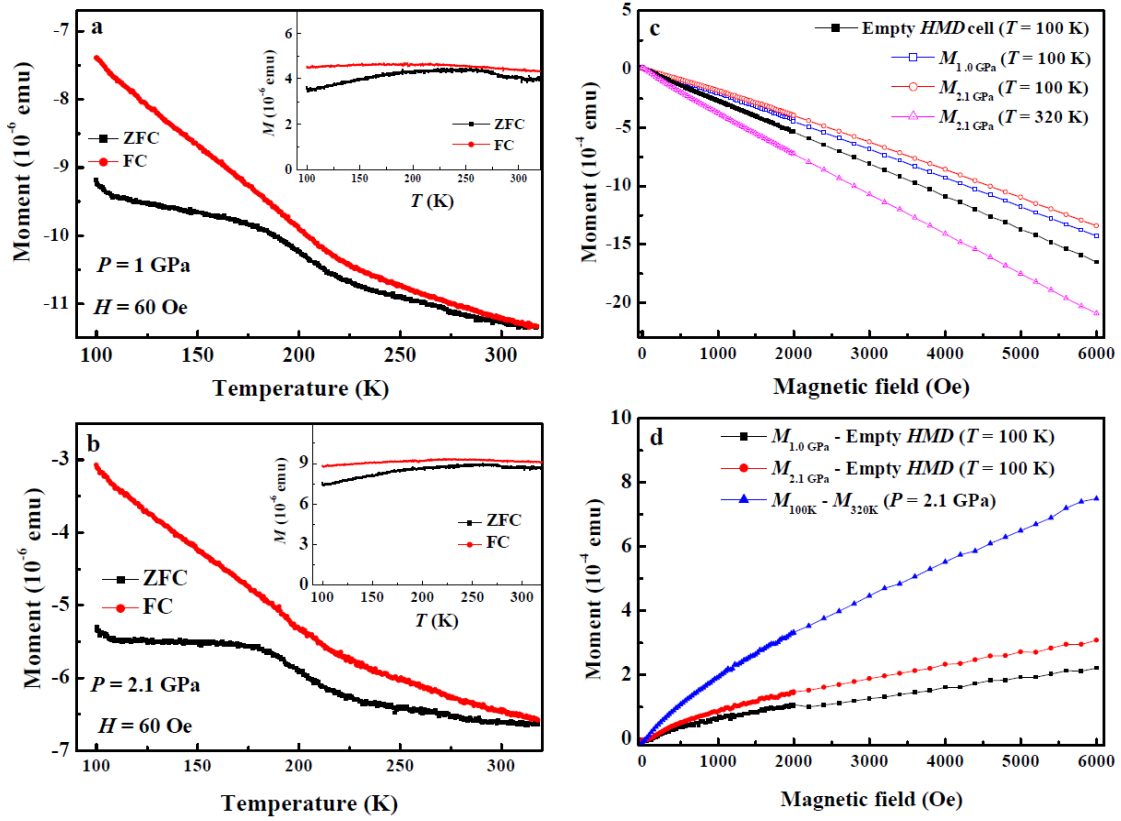

**Fig. 4 | Magnetic properties for LuH$_{2\pm x}$N$_y$ at different pressures. a, b** Temperature dependence of magnetic moment for LuH$_{2\pm x}$N$_y$ under pressures of 1 GPa and 2.1 GPa, respectively. Shown in the main panels are raw data. The insets show the corresponding magnetization measured in ZFC and FC modes with the background subtracted. **c** *M(H)* curves at 100 K under pressures of 1.0 GPa (open square) and 2.1 GPa (circle), and one curve at 320 K under 2.1 GPa (up triangle), respectively. The *M*(H) curve measured at 100 K for the empty HMD cell is also shown here (solid square). **d** *M(H)* curves with different background signals removed.

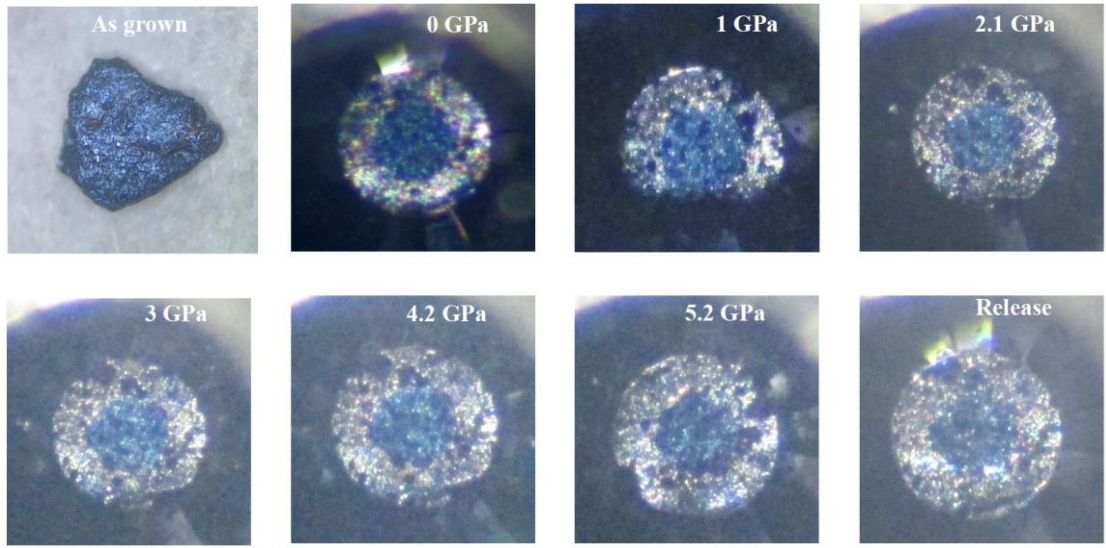

**Fig. 5 | The optical microscope images of LuH$_{2\pm x}$N$_y$ at different pressures.**

# Supplementary Information

# Absence of near-ambient superconductivity in LuH$_{2\pm x}$N$_y$

Xue Ming[†], Ying-Jie Zhang[†], Xiyu Zhu[*], Qing Li[*], Chengping He, Yuecong Liu, Bo Zheng, Huan Yang, and Hai-Hu Wen[*]

National Laboratory of Solid State Microstructures and Department of Physics, Collaborative Innovation Center of Advanced Microstructures, Nanjing University, Nanjing 210093, China

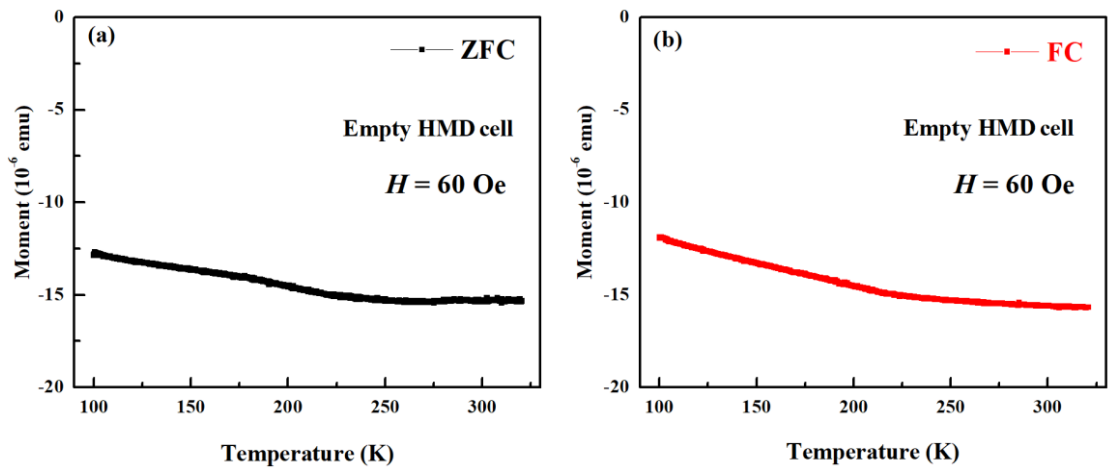

**Extended Fig. 1 | Temperature dependent magnetic moment of the empty HMD cell with the applied magnetic field of $H$ = 60 Oe.** The **a** ZFC and **b** FC curves are used as the background signals to obtain the pure magnetic moment of the sample in d.c. magnetization measurements.

# LuH₂

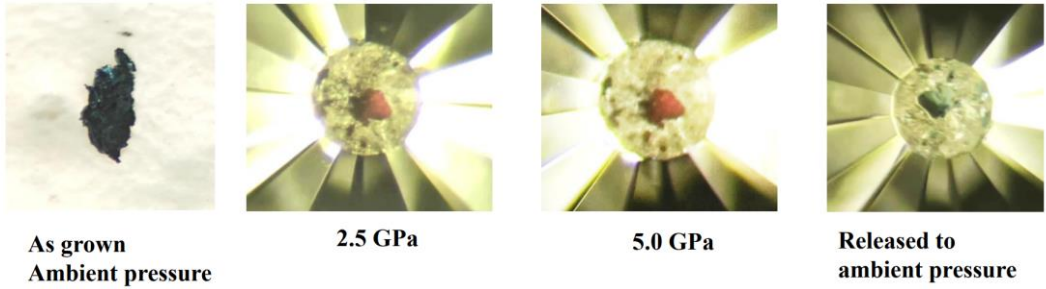

As grown
Ambient pressure | 2.5 GPa | 5.0 GPa | Released to ambient pressure

**Extended Fig.2 | Color change of LuH₂.** The sample LuH$_2$ at ambient pressure shows a dark-blue color, it shows pink and red color at pressures of 2.5 GPa or 5.0 GPa. The color changes to a light-blue when the pressure is released.